\documentclass[conference]{IEEEtran}
\IEEEoverridecommandlockouts

\usepackage{algorithmic}
\usepackage{graphicx}
\usepackage{textcomp}
\usepackage{xcolor}
\usepackage{xspace}
\usepackage{graphicx}
\usepackage[displaymath, mathlines]{lineno}
\usepackage{ifthen}
\usepackage[normalem]{ulem} 
\usepackage{xcolor,colortbl}
\usepackage{hyperref}

\newboolean{showedits}
\setboolean{showedits}{true} 
\ifthenelse{\boolean{showedits}}
{
	\newcommand{\del}[1]{\textcolor{red}{\sout{#1}}} 
	\newcommand{\nbe}[3]{
		{\colorbox{#3}{\bfseries\sffamily\scriptsize\textcolor{white}{#1}}}
		{\textcolor{#3}{\sf\small$\blacktriangleright$\textit{#2}$\blacktriangleleft$}}}
}{
	\newcommand{\del}[1]{} 
	
	\newcommand{\nbe}[3]{}
}

\newboolean{showcomments}
\setboolean{showcomments}{true} 
\newcommand{\id}[1]{$-$Id: scgPaper.tex 32478 2010-04-29 09:11:32Z oscar $-$}

\ifthenelse{\boolean{showcomments}}
{
	\newcommand{\nbc}[3]{
		{\colorbox{#3}{\bfseries\sffamily\scriptsize\textcolor{white}{#1}}}
		{\textcolor{#3}{\sf\small$\blacktriangleright$\textit{#2}$\blacktriangleleft$}}}
	
}{
	\newcommand{\nbc}[3]{}
	
}

\usepackage[most]{tcolorbox}
\ifthenelse{\boolean{showedits}}
{
	\newtcolorbox{inserted}{%
		title=Inserted text:,
		colframe=blue,colback=blue!5!white,
		breakable,
		leftrule=0mm, 
		bottomrule=0mm,
		rightrule=0mm,
		toprule=0mm,
		arc=0mm, outer arc=0mm,
		oversize
	}
	\newtcolorbox{deleted}{%
		title=Deleted text:,
		colframe=red,colback=red!5!white,
		breakable,
		leftrule=0mm, 
		bottomrule=0mm,
		rightrule=0mm,
		toprule=0mm,
		arc=0mm, outer arc=0mm,
		oversize
	}
	\newtcolorbox{refactored}{%
		title=Rewritten text:,
		colframe=blue,colback=red!5!white,
		breakable,
		leftrule=0mm, 
		bottomrule=0mm,
		rightrule=0mm,
		toprule=0mm,
		arc=0mm, outer arc=0mm,
		oversize
	}
}{

}
\newboolean{isblinded}
\setboolean{isblinded}{true}
\ifthenelse{\boolean{isblinded}}
{\newcommand\blind[1]{BLINDED\xspace}}
{\newcommand\blind[1]{#1\xspace}}

\newcommand{\commented}[1]{}

\newcommand{\eg}{\emph{e.g.,}\xspace}
\newcommand{\ie}{\emph{i.e.,}\xspace}

\newcommand{\ignore}[1]{}
\newcommand\tool{NASRA\xspace}

\newcommand{\keyword}[1]{{\code{\color{blue} #1}}}
\newcommand{\literal}[1]{{\code{\color{purple} #1}}}
\newcommand{\gkeyword}[1]{{\code{\color{violet} #1}}}
\newcommand{\type}[1]{\code{{\color{green!20!black} #1}}}

\newcommand{\boxit}[1]{\vspace{0.2cm}
\noindent
\fbox{
\begin{minipage}{8.3cm}
\small #1 
\end{minipage}
}
}

\newcommand{\code}[1]{\texttt{#1}}

\definecolor{source}{gray}{0.9}
\usepackage[utf8]{inputenc}
\usepackage{url}
\usepackage{booktabs}
\usepackage{longtable}
\usepackage{xspace}
\usepackage[most]{tcolorbox}

\usepackage{multirow}

\begin{document}

\title{Naturalistic Static Program Analysis}


\author{
\IEEEauthorblockN{Mohammad Mehdi Pourhashem Kallehbasti}
\IEEEauthorblockA{Department of Electrical and Computer Engineering\\University of Science and Technology of Mazandaran\\P.O. Box 48518-78195, Behshahr, Iran\\pourhashem@mazust.ac.ir}
\and
\IEEEauthorblockN{Mohammad Ghafari}
\IEEEauthorblockA{TU Clausthal, Germany\\mohammad.ghafari@tu-clausthal.de}}

\maketitle

\begin{abstract}

Static program analysis development is a non-trivial and time-consuming task.
We present a framework through which developers can define static program analyses in natural language.
We show the application of this framework to identify cryptography misuses in Java programs, and we discuss how it facilitates static program analysis development for developers.
\end{abstract}



\section{Introduction}


Static program analysis is the art of examining programs without requiring to execute the code.
However, 
static analysis tools generate false positives and tuning them requires expertise.
Likewise, program analysis development requires a deep knowledge of compiler or mastering an analysis framework.

End-user programming is a set of techniques that enable end users to write programs at a level of complexity that is adequate to their practices, background, and skills.
For instance, it includes visual languages to program robots through visual blocks~\cite{BARAKOVA2013}, and simplified programming languages to translate English sentences to Bash commands~\cite{Victoria2018}.
We believe that end-user programming techniques can also help to hide the complexity of writing a static program analysis task for non-professional programmers and empower them in this domain. 

We introduce \tool (NAturalistic Static pRogram Analysis), a framework that enables developers to define a program analysis task in natural language (NL), and it generates the corresponding Query Language (QL) query that underlies CodeQL program analysis engine.\footnote{https://codeql.github.com}
We illustrate the application of this framework to find cryptography misuses in Java programs.
\tool is open source and publicly available.\footnote{https://doi.org/10.5281/zenodo.7495044}


%
%
%
%

The ultimate goal of \tool is to enable ``naturalistic'' static program analysis development in a way that developers can specify what they need without deep knowledge of static program analysis and how a specific framework works.
Its higher level of abstraction than existing static analysis frameworks may facilitate a more intuitive formulation of program analysis tasks.
Similarly, its agnostic nature to programming languages can provide a cross-language interface for program analysis, which obviates the need to learn the specifics of a program analysis framework.
This paper presents a preliminary step to realize the above goal.



\section{The \tool Framework}
\label{sec:motivation}

Cryptography is an essential component to security, but it is one of the notorious topics where developers struggle a lot~\cite{Hazhirpasand2021,Hazhirpasand2020}.
%
Locating the \literal{init} method invoked on a \literal{Cipher} object is often deemed to be the first step to analyze cryptography code in Java programs.
For instance, in CodeQL, one should write the following query to implement this task.

%

\boxit{
		\keyword{from } \type{MethodAccess }\literal{init}\\
		\keyword{where} \literal{init}\type{.getMethod().getName()} = \literal{"init"} \keyword{and}\\ \literal{init}\type{.getReceiverType().getName()} = \literal{"Cipher"} 
\\
		\keyword{select} \literal{init}
}


We have developed a framework, called \tool, that enables a more intuitive formulation of the above task in the form below:

\boxit{
An object of \literal{Cipher} invokes \literal{init}.
}




\tool is a rule-driven synthesizer.
We rely on predefined rules due to a lack of trustworthy labeled examples required for a data-driven approach in this domain.
\tool receives a program analysis inquiry in natural language, applies semantic parsing, and generates CodeQL commands.
The input inquiry should comply with a subset of the syntax of Attempto Controlled English (ACE) controlled natural language. 
We use Attempto Parsing Engine (APE), a tool that receives a series of ACE statements and produces the corresponding Discourse Representation Structures (DRS) that is a semantic representation of the input text.
\tool applies the translation rules, explained later in this section, on the given DRS and produces the corresponding CodeQL statements. 
Thanks to APE, the way one can formulate \tool statements is very flexible and there is no need for absolute correspondence with the \tool syntax.
We chose CodeQL as our code analysis engine because it is an industry-leading and community-powered tool, and its publicly available to all GitHub users without any installation hassle.
%
%
%
To employ \tool for a new static analysis framework, only the transformation rules have to be adapted. To support a new application domain, we should identify the types of queries that the current syntax does not support, add the corresponding production rules to the syntax, and develop translations for them.
\tool is open source, and currently, supports program analysis tasks that concern cryptography misuses in Java programs.

\subsection{Syntax and Semantics}

Each \tool query comprises one or more  Statement. The syntax is shown below (terminals have different color).
%

\boxit{
	$
	Query ::= Statement \;Query \;|\; Statement\\
	Statement ::= BasicStatement \;|\; LogicalStatement\\
	\;|\; Extension\\
	BasicStatement ::= Exp \;\gkeyword{is} \; (Exp \;|\; \gkeyword{in} \;List)\\
	Exp ::= Prefix \;Exp \;|\; \texttt{type} \;|\;ID \;|\; Literal\\
	Prefix ::= ((adjective | \varepsilon) \;attribute\; \textrm{\gkeyword{of}})\\
	LogicalStatement ::= Statement \;\textrm{\gkeyword{and}} \;Statement\;|\;\\
	Statement \;\textrm{\gkeyword{or}} \;Statement\;|\; \textrm{\gkeyword{It is false that}} \;Statement\\
	\;|\; \textrm{\gkeyword{If}} \;Statement \;\textrm{\gkeyword{then}} \;Statement$
}

\paragraph{Expression}
The smallest building block is \textbf{Exp}. 
It includes a Literal (String or int) or an ID (user defined identifier) that are directly mapped to CodeQL expressions. 
%
%
An Exp can also be a CodeQL type such as 
class, variable, and method access that are mapped to \type{Class}, \type{Variable}, and \type{MethodAccess}, respectively.

\paragraph{Prefix}
Each Exp can have an optional \textbf{Prefix} in the form of \emph{``attribute of''} that indicates an attribute of the expression.
For instance, name, type, argument, and method are attributes of an entity (\ie Exp), and they correspond to \type{\code{getName()}}, \type{\code{getType()}}, \type{\code{getArgument()}}, and \type{\code{getMethod()}} methods in CodeQL, respectively.

For example, ``name of \literal{method1}'' is an Exp, where ``name'' is an attribute and \literal{method1} is an ID, and the whole expression is translated to \texttt{\literal{method1}\type{.getName()}} in CodeQL.

Additionally, the attribute itself can have an optional ordinal number as an \emph{adjective}, like \textit{second} in the Exp ``second argument of \literal{init}'' that is translated to ``\texttt{\literal{init}\type{.getArgument(1)}}'', where second is translated to 1 as an argument according to zero-based numbering.

Note that ``attribute of'' can be repeated several times, where each attribute may have an adjective. For example, the Exp ``The type of the second argument of \literal{init}'' has one ID (\ie \literal{init}) and two attributes (\ie type and argument).

\paragraph{Basic Statement}
Each \textbf{BasicStatement} is a statement that can serve as a Boolean condition as well as an assumption. 

As a \emph{Boolean condition}, BasicStatement produces equivalence of two Exps, as well as membership of an Exp in a list.
In ``Exp is Exp'' structure, both sides of equivalence are Exps and they need to be equal, while in ``Exp is in List'' structure, the Exp needs to be equal to an item in a list. 
Accordingly, a statement like ``\literal{arg1} is in [\literal{"RSA"}, \literal{"AES"}].'' is a disjunctive expression and can be rephrased to ``\literal{arg1} is \literal{"RSA"} or \literal{arg1} is \literal{"AES"}.'', that is ultimately translated to ``\texttt{\literal{arg1} = \literal{"RSA"} \keyword{or} \literal{arg1} = \literal{"AES"}}''.

The syntax structure \textit{Exp is Exp} can also produce \emph{assumptions} when the second Exp is a CodeQL \texttt{type}.
The assumptions are mapped to the \keyword{from} part of a CodeQL query. 
For instance, the statement ``\literal{var1} is a variable.'' translates to ``\texttt{\type{Variable} \literal{var1}}'' and belongs to the \texttt{\keyword{from}} part.

\paragraph{Logical Statement}


A \textbf{LogicalStatement} can be a
negation, conjunction, disjunction, or implication. 
For example, ``If \literal{arg1} is \literal{"RSA"} then \literal{arg2} is \literal{"AES"}.'' is translated to ``\texttt{\keyword{not} (\literal{arg1} = \literal{"RSA"}) \keyword{or} \literal{arg2} = \literal{"AES"}}'' in CodeQL, since $p\Rightarrow q$ is equivalent to $\lnot p \vee q$.

\subsection{Extensibility}

One can extend \tool to cover auxiliary statements and statement patterns.
Their corresponding production rules are as follows.

\boxit{
$Extension ::= Pattern \;|\; AuxiliaryStatement$
}

We introduce these features through 
three statement patterns and one auxiliary statement that are helpful to cover constraints on using Java cryptography objects.

\subsubsection{Patterns}

We present three patterns that extend \textbf{Pattern} nonterminal in the syntax.
We discuss each in the following.

\paragraph{Invocation}

We use this pattern to state that a method is \emph{invoked by} an instance of a specific class. 
It can also be used to make sure that there is \emph{no invocation} of a method by any instance of a specific class.

\boxit{
	$
	Pattern_1::=\gkeyword{\textrm{An object of }}ID \;(\gkeyword{\textrm{invokes}}|\gkeyword{\textrm{does not invoke}})\;ID.$}







The \tool query shown in Section~\ref{sec:motivation} is an example of this pattern.
The transformation follows a number of steps.
First, 
a \type{MethodAccess} is declared with the same name used in the \tool statement (\ie \literal{init}). 
Then the conditions need to be added to the \keyword{where} part.
Specifically, the name of the method of the \type{MethodAccess} \literal{init} should be \literal{"init"} that is stated in the second line.
Finally, a \type{MethodAccess} has a receiver, that is the object invoking its method.
In this case, the name of the type of the receiver should be \literal{"Cipher"}, that is expressed in CodeQL in the third line.

If one needs to make sure that \emph{no invocation} occurs, an existential quantifier must be used, as shown in the following.

\boxit{
	\keyword{from}\\
		\keyword{where not} (\keyword{exists} (\type{MethodAccess} \literal{init}
        $|$
		\literal{init}.\type{getMethod().getName()} = \literal{"init"}
		\keyword{and} \\ \literal{init}.\type{getReceiverType().getName()} = \literal{"Cipher"}))
}
		
It means that there is no such \type{MethodAccess} \literal{init} that has these conditions.
We can state this in \tool in the form below.
 
\boxit{
An object of \literal{Cipher} doesn't invoke \literal{init}.
}

\paragraph{Partial order constraints}

This pattern enables one to put \emph{partial order} constraints on method invocations. 
In other words, one can enforce a method invocation to be preceded (or followed) by another method invocation.

\boxit{
	$
	\!\!Pattern_2\!\!::=\!\!\textrm{MethodName (\gkeyword{precedes}}| \textrm{\gkeyword{follows}) MethodName.}$}

For example, there are two steps in CodeQL for stating that ``invocation of \literal{getInstance} is earlier than invocation of \literal{init}".
%
First, one should specify that both methods are in the same scope.
Next, the line number of the preceding method invocation has to be smaller than the line number of the other method invocation.
This is shown below.


\boxit{
\literal{getInstance}\type{.getEnclosingCallable()} = \literal{init}\type{.getEnclosingCallable()}
\keyword{and} \\ \literal{getInstance}\type{.getLocation().getEndLine()} \texttt{<} \literal{init}\type{.getLocation().getEndLine()}
}

We can express this query in \tool as follows.

\boxit{
\literal{getInstance} precedes \literal{init}.
}




\paragraph{Method signature constraint}
It is possible to express signature of a method using \textbf{Pattern$_3$}.

\boxit{
	$Pattern_3 ::= \textrm{MethodName's\,\gkeyword{signature is} }  List.
	$
}

 A method signature can be seen as an ordered list of data types.
This list contains names of data types as strings, such that the first string is the name of the first argument's data type and so on. For example, the following \tool query states that \literal{getInstance} method has two arguments and the names of their types are \literal{"int"} and \literal{"Certificate"}, respectively.

\boxit{
\literal{getInstance}'s signature is \literal{["int", "Certificate"]}.}

 This query is translated to the following CodeQL query.
 
\boxit{
(\keyword{count} (\literal{getInstance}\type{.getAnArgument()}) = \literal{2}) \keyword{and} \literal{getInstance}\type{.getArgument(0).getType().\\toString()}=\literal{"int"} \keyword{and} \literal{getInstance.}\\\type{getArgument(1).getType().toString()}= \literal{"Certificate"}}

First, the number of arguments is set to the size of the user defined list, then the type of arguments are constrained one by one. \texttt{\type{count (method.getAnArgument())}} returns the number of arguments of the method. \type{getArgument(i)} returns the argument number $i$ in the given method, \type{getType()} returns the type of the given argument, and finally \type{toString()} converts the given data type to a String.

\subsubsection{AuxiliaryStatement}

We aim to find misuses in code that violate one or more mandatory constraints.
For instance, 
suppose that if the second argument of \literal{init} method is \literal{"private key"} then it is mandatory that the encryption algorithm, \ie the second argument of \literal{getInstance} method is \literal{"RSA"}, 
and also if the encryption algorithm is \literal{"AES"} then it is mandatory that the mode of encryption, \ie the first argument of the \literal{getInstance} method, is \literal{"CBC"}. 
The following \tool query will find such violations.

\boxit{
It is false that if the type of the second argument of \literal{init} is \literal{"PrivateKey"}, then the algorithm of \literal{getInstance}'s first argument is \literal{"RSA"} or it is false that if the algorithm of \literal{getInstance}'s first argument is \literal{"AES"} then the mode of \literal{getInstance}'s first argument is \literal{"CBC"}.}


In order to find any violation of these constraints, disjunction of their negation has to be stated in the query.\footnote{
For example, in ``X is driving in an urban area($Cond_1$). It is necessary that X is driving slower than 60 km/h ($Cons_1$). It is necessary that X fastens the seat belt ($Cons_2$).", the query needs to find an X that is driving in an urban area and is driving faster than 60 km/h or is not using the seat belt. If we assign a Boolean variable to each statement as mentioned in the statements, it should aim $Cond_1 \wedge (\lnot Cons_1\vee \lnot Cons_2)$ whose necessity part is translated to the disjunction of negation of two constraints.
} 
Nevertheless, the above statement becomes much longer and harder to comprehend as the number of constraints increases.

We define \emph{auxiliary statements} to ease the formulation as well as the comprehension of complex queries for developers.
Particularly, 
\textbf{NecessityStatements} are auxiliary statements that enable developers to enforce mandatory constraints in short and independent statements.
It starts with ``\emph{It is necessary that}" and follows the syntax below. 

\boxit{
$
NecessityStatement ::= \textrm{\gkeyword{It is necessary that}} \; Statement.
$
}

Therefore, instead of writing disjunction of negation of all constraints in one single statement, developers can benefit this construct (i.e., NecessiyStatement) to define all such constraints in several statements within a query.
Accordingly, the single but long previous statement can be stated as two separate statements shown below.

\boxit{
It is necessary that if the type of the second argument of \literal{init} is \literal{"PrivateKey"}, then the algorithm of \literal{getInstance}'s first argument is \literal{"RSA"}.\\
It is necessary that if the algorithm of \literal{getInstance}'s first argument is \literal{"AES"} then the mode of \literal{getInstance}'s first argument is \literal{"CBC"}.
}

Necessity statements are treated differently from other statements. 
If there is only one NecessityStatement, its enclosing statement is negated and added to the \keyword{where} part of the CodeQL query. 
If there are more than one, 
\eg $n$ constraints $Cons_1$, $Cons_2$, ..., $Cons_n$, then the ``\code{(\keyword{not} $TCons_1$ \keyword{or} \keyword{not} $TCons_2$ \keyword{or} ... \keyword{or} \keyword{not} $TCons_n$)}'' will be added to the \texttt{where} part, where $TCons_i$ is the translation of $Cons_i$.




\section{Working Examples}
\label{sec:casestudy}

%
Cipher is one of the most misused APIs in Java cryptography~\cite{Hazhirpasand2020}. 
Listing~\ref{cipherexample} shows how to create a \literal{Cipher} object in Java.
We should call the \literal{Cipher}'s \literal{getInstance} method.
This method receives a number of arguments.
The first one is \texttt{transformation} that is a string containing three parts separated by ``\texttt{/}". These parts are \texttt{algorithm}, \texttt{mode}, and \texttt{padding}, respectively. 
Next, we should call the \literal{init} method on the \literal{cipher} object with two arguments to indicate the operation mode of the \literal{cipher}, and to initialize this object with a Key or Certificate.

\begin{lstlisting}[numbers=none, label=cipherexample, caption=Setting up the Cipher object in Java]
Cipher cipher = Cipher.getInstance("AES/ECB/PKCS5Padding");
cipher.init(Cipher.ENCRYPT_MODE,new SecretKeySpec(keyBytes, "AES"));
\end{lstlisting}

In the rest of this section, we present three different program analysis tasks to ensure secure use of Java Cipher. 

\subsection{Key vs. Algorithm}
\label{subsec:casestudy_01}

\emph{
\textbf{Task 1:}
If the key has a type of PublicKey, PrivateKey, or Certificate, or encryption mode is WRAP\_MODE  or UNWRAP\_MODE, then algorithm of transformation must be ``RSA''.
}

\newpage
Listing~\ref{keyvsalgorithm}
shows how to check this constraint in CodeQL.

\begin{lstlisting}[numbers=none, label={keyvsalgorithm}, caption= Key vs. Algorithm constraint in CodeQL]
from MethodAccess getInstance, MethodAccess init
where init.getMethod().getName() = "init" and init.getReceiverType().getName() = "Cipher" and getInstance.getMethod().getName() = "getInstance" and getInstance.getReceiverType().getName() = "Cipher" and (((init.getArgument(0).toString() = "Cipher.WRAP MODE" or init.getArgument(0).toString() = "Cipher.UNWRAP MODE") or (init.getArgument(1).getType().toString() = "java.security.PublicKey" or init.getArgument(1).getType().toString() = "java.security.PrivateKey" or init.getArgument(1).toString() = "java.security.cert.Certificate")) and not(getInstance.getArgument(0).toString().replaceAll("\","").splitAt("/",0) = "RSA"))
select getInstance, init
\end{lstlisting}

This constraint can be expressed in \tool as follows.

\boxit{
An object of \literal{Cipher} invokes \literal{init}. An object of \literal{Cipher} invokes \literal{getInstance}. It is necessary that if \literal{init}'s first argument is in 
	{[}\literal{"Cipher.WRAP\_MODE"}, \literal{"Cipher.UNWRAP\_MODE"}{]} or the type of the second argument of \literal{init} is in 
	{[}\literal{"PublicKey"}, \literal{"PrivateKey"}, 
	\literal{"Certificate"}{]} then 
	the algorithm of \literal{getInstance}'s first argument is \literal{"RSA"}.
}


\subsection{Algorithm vs. Transformation Mode}
\label{subsec:casestudy_02}

\emph{
\textbf{Task 2:}
If the algorithm of transformation is ``RSA'' then the mode of transformation must be either ``'' or  ``ECB''.
}

Listing~\ref{algorithmvsmode}
shows the corresponding query to check this constraint in CodeQL.
We should look for code in which the algorithm is ``RSA'', but neither ``ECB'' nor ``'' is set for the mode. 

\begin{lstlisting}[numbers=none, label={algorithmvsmode}, caption= Algorithm vs.  Transformation Mode constraint in CodeQL]
from MethodAccess getInstance
where getInstance.getMethod().getName() = "getInstance" and getInstance.getReceiverType().getName() = "Cipher" and (getInstance.getArgument(0).toString().replaceAll("\"","").splitAt("/", 0) = "RSA") and not (getInstance.getArgument(0).toString().replaceAll("\"","").splitAt("/", 1) = "" or getInstance.getArgument(0).toString().replaceAll("\"","").splitAt("/", 1) = "ECB")
select getInstance
\end{lstlisting}

This constraint can be expressed in \tool as follows.

\boxit{
	An object of \literal{Cipher} invokes \literal{getInstance}. It is necessary that if the algorithm of \literal{getInstance}'s first argument is \literal{"RSA"} then the mode of \literal{getInstance}'s first argument is in [\literal{""}, \literal{"ECB"}].
}

Thanks to Attempto Parsing Engine (APE), \tool statements do not need to exactly follow the syntax rules meaning that a degree of freedom in paraphrasing is possible.
For instance, the part ``the algorithm of \literal{getInstance}'s first argument is \literal{"RSA"}'' can also be written in two other forms:

\boxit{
(i) the algorithm of the first argument of \literal{getInstance} is \literal{"RSA"}.\\
(ii) \literal{"RSA"} is the algorithm of \literal{getInstance}'s first argument.
}

\subsection{Transformation and Encryption Mode vs. Signature}
\label{subsec:casestudy_03}

\emph{
\textbf{Task 3:}
If the transformation mode is either of ``CBC'', ``PCBC'', ``CTR'', ``CTS'', ``CFB'', or ``OFB'', and the encryption mode is not ``Cipher.ENCRYPT\_MODE'', then the invoked init method should not have any of the following signature:
init(encmode, cert), init(encmode, cert, ranGen), init(encmode, key), init(encmode, key, ranGen).
}

Listing~\ref{modevssignature} presents how to enforce this constraint in CodeQL.

\begin{lstlisting}[numbers=none, label={modevssignature}, caption=Transformation and Encryption mode vs. Signature constraint in CodeQL]
from MethodAccess getInstance, MethodAccess init 
where init.getMethod().getName() = "init" and init.getReceiverType().getName() = "Cipher" and getInstance.getMethod().getName() = "getInstance" and getInstance.getReceiverType().getName() = "Cipher" and ((getInstance.getArgument(0).toString().replaceAll("\"","").splitAt("/", 1) = "CBC" or getInstance.getArgument(0).toString().replaceAll("\"","").splitAt("/", 1) = "PCBC" or getInstance.getArgument(0).toString().replaceAll("\"","").splitAt("/", 1) = "CTR" and getInstance.getArgument(0).toString().replaceAll("\"","").splitAt("/", 1) = "CTS" or getInstance.getArgument(0).toString().replaceAll("\"","").splitAt("/", 1) = "CFB" or getInstance.getArgument(0).toString().replaceAll("\"","").splitAt("/", 1) = "OFB") and not (init.getArgument(0).toString() = "Cipher.ENCRYPT_MODE")) and ((count (getInstance.getAnArgument()) = 2 and getInstance.getArgument(0).getType().toString() = "int" and getInstance.getArgument(1).getType().toString() = "Certificate") or (count (getInstance.getAnArgument()) = 3 and getInstance.getArgument(0).getType().toString() = "int" and getInstance.getArgument(1).getType().toString() = "Certificate" and getInstance.getArgument(2).getType().toString() = "SecureRandom") or (count (getInstance.getAnArgument()) = 2 and getInstance.getArgument(0).getType().toString() = "int" and getInstance.getArgument(1).getType().toString() = "Key") or (count (getInstance.getAnArgument()) = 3 and getInstance.getArgument(0).getType().toString() = "int" and getInstance.getArgument(1).getType().toString() = "Key" and getInstance.getArgument(2).getType().toString() = "SecureRandom"))
select init, getInstance
\end{lstlisting}

The implementation of this task in \tool is shown below.

\boxit
{
An object of \literal{Cipher} invokes \literal{getInstance}. An object of \literal{Cipher} invokes \literal{init}. 
	It is necessary that if the mode of \literal{getInstance}'s first argument is in \literal{[\literal{"CBC"},\literal{"PCBC"},"CTR","CTS","CFB","OFB"]} and \literal{init}'s first argument is not \literal{"Cipher.ENCRYPT\_MODE"} then \literal{getInstance}'s signature is not \literal{["int","Certificate"]} and is not \literal{["int","Certificate","SecureRandom"]} and is not \literal{["int","Key"]} and is not \literal{["int","Key","SecureRandom"]}.
}

\subsection{Discussion}

%
Table~\ref{Table:comparison} presents the number of \emph{distinct} operators and operands (\ie vocabulary), and the total number of operators and operands (\ie length) needed for each analysis task.\footnote{In \tool, we consider user defined terminals such as init, ``RSA'', and getInstance as operands and count the rest of language constructs as operators.}
Evidently, queries in \tool are significantly shorter than queries in CodeQL (\ie up to 87\% reduction in length), and they consume a lot fewer programming constructs (\ie up to 38\% fewer vocabularies).
We computed Halstead complexity measures to estimate the coding time and the difficulty to write or understand these queries~\cite{halstead1977elements}.
The results showed that developers require a lot less effort and time to develop these tasks in \tool than in CodeQL.
%

%
We also asked ten developers to share their opinion about queries in \tool. 
They unanimously stated that they are succinct and easy to understand, and one commented that ``\emph{these queries read like API documentation}".

It is noteworthy that \tool's performance, \ie how well it can detect API misuses, depends on its underlying analysis framework which is currently CodeQL.
In other words, \tool obviates the low-level details needed to define static program analyses, but the issues with false positives remain to be relevant.
Moreover, despite being natural, the use of \tool still requires knowledge of its syntax.

\begin{table}
\caption{CodeQL vs. \tool}
\begin{tabular}{|l|ll|ll|}
\hline
\multicolumn{1}{|c|}{\multirow{2}{*}{\textbf{Analysis Task}}}    & \multicolumn{2}{c|}{\textbf{Vocabulary}}           & \multicolumn{2}{c|}{\textbf{Length}}               \\ \cline{2-5} 
\multicolumn{1}{|c|}{}                                           & \multicolumn{1}{l|}{CodeQL$\!\!\!$} & \tool$\!\!\!$ & \multicolumn{1}{l|}{CodeQL$\!\!\!$} & \tool$\!\!\!$ \\ \hline
Key vs. Algorithm (\ref{subsec:casestudy_01})  & \multicolumn{1}{l|}{32}     & 19                   & \multicolumn{1}{l|}{179}    & 39                   \\ \hline
Algorithm vs. Mode (\ref{subsec:casestudy_02}) & \multicolumn{1}{l|}{27}     & 18                   & \multicolumn{1}{l|}{107}    & 24                   \\ \hline
Mode vs. Signature (\ref{subsec:casestudy_03}) & \multicolumn{1}{l|}{42}     & 26                   & \multicolumn{1}{l|}{434}    & 56                   \\ \hline
\end{tabular}
\label{Table:comparison}
\end{table}

\section{Related work}
\label{sec:literature}

Mapping a natural language statement into a formal representation has received great attention in the community but not much in the program analysis development domain.


Schlegel et al. developed an end-user programming paradigm for Python, that maps natural language commands into Python code~\cite{Schlegel2019}.
%
Landhauber et al. proposed a domain agnostic command interpreter that receives natural language commands in English and uses ontology to produce relevant API calls~\cite{NLCI2017}.
%
%
Yaghmazadeh et al. developed SQLIZER, a system to automatically synthesize SQL queries from a natural language~\cite{Yaghmazadeh2017}.
Luo et al. investigated the translation from a natural language query to visualization with the goal of simplifying the creation of data visualizations~\cite{Luo2022}.
%

Heyman et al. developed a Python code completion tool that enriches developers' code with the natural language description of the intended data science task~\cite{Heyman2021}.
%
%
Nguyen et al. presented an approach that takes as input an English description of a programming task and synthesizes the corresponding API code template for the task~\cite{Nguyen2018}.
Desai et al. built a general framework for constructing program synthesizers that take natural language inputs and produce expressions in a target Domain Specific Language~\cite{Desai2016}.
Zhai et al. proposed a search-based technique to automatically translate NL comments to formal program specifications that specify the expected pre and post conditions~\cite{Zhai2020}.
\newpage

The work presented in this paper is also related to cryptography domain. 
There exist tools that find cryptography misuses~\cite{Zhang22} and libraries that facilitate the adoption of cryptography for developers~\cite{Kafader21}. Nevertheless, none of them employed a natural language approach.





\section{Conclusion}
\label{sec:end}

We introduced \tool, an open-source framework to define static program analyses in natural language.
%
We demonstrated the application of this framework to find misuses in Java cryptography.
The ultimate goal of \tool is to enable a naturalistic way to develop static program analyses, which is usable for mainstream developers. To realize this goal, further studies are needed to determine \tool's effectiveness in real-world settings.
%
The expressiveness of its queries and the effort required to extend it to other problem domains have to be investigated as well.
Finally, automatic translation without pre-defined rules is also an exciting future research direction.

\bibliographystyle{IEEEtran}
\bibliography{reference.bib}




%

\end{document}